# Space Environment and Debris: A Review of Micro-Meteoroids and Orbital Debris Impact Protection

Binkal Kumar Sharma[a], Harshitha Baskar[b]

[a]*University of Bremen, Germany*
[b]*Independent Researcher, India*

**Abstract**

The threat posed by MMOD to spacecraft has escalated with the growing density of orbital objects, driven by the proliferation of satellite constellations such as Starlink and OneWeb. This paper reviews the current challenges and advancements in MMOD impact protection, emphasizing innovations in shielding technologies, material science, and sustainability practices. The synthesis of recent developments highlights the role of hybrid materials, additive manufacturing, and international collaboration in ensuring spacecraft resilience while promoting orbital sustainability.

**Introduction**

The establishment of sustainable habitats on extraterrestrial bodies has long been a key objective in human space exploration, driven by both scientific curiosity and the strategic imperative for long-term survival beyond Earth [1, 2]. This ambition gained significant momentum on 20 July 1969, when the Apollo 11 mission successfully landed the first humans on the Moon [3]. Since that historic milestone, advances in space technology have increasingly focused on overcoming multifaceted challenges, including harsh radiation environments, extreme temperature fluctuations, and the persistent threat of micrometeoroids and orbital debris (MMOD) impacts [4, 5, 6, 7]. Spacecraft operating in Low Earth Orbit (LEO) and beyond are continuously exposed to hypervelocity impacts (often exceeding 7 km/s) that can compromise critical components [8, 9, 6, 7, 10]. In addition, the rapid proliferation of satellites, exemplified by recent mega-constellation initiatives by several private companies such as SpaceX, and the filing of nearly 330,000 satellites by emerging nations [11, 12, 13], have heightened concerns over orbital congestion and environmental degradation in space [2, 14]. Furthermore, the likelihood of MMOD impacts on spacecraft has increased significantly. [15, 13]. This thesis underscores the critical necessity of incorporating protective measures against MMOD hazards. It systematically evaluates the potential of composite materials as a viable and effective solution for enhancing spacecraft resilience against MMOD-induced impacts.

Traditional materials, such as metals, metal-alloys and polymers, although fundamental to spacecraft design, often become inadequate under such extreme conditions, with metals facing mass penalties and brittleness, and polymers suffering degradation under repetitive high-energy collisions [4, 9]. These challenges have spurred the development of advanced shielding systems, including multi-layered Whipple shields that employ a sacrificial outer layer to fragment and dissipate incoming debris, thus protecting the core structure[16, 6, 17]. Recent research in material science also point to hybrid metal–polymer composites as a promising solution, integrating the high tensile strength of metals with the lightweight, energy-absorbing attributes of polymers to achieve a superior balance of performance and mass efficiency [18, 19]. The selection and processing of spacecraft materials to ensure durability in the harsh space environment are guided by NASA's material standards such as NASA-Std-6016B. [20].This review paper explores the different types of MMOD protection techniques, and how metal polymer composite can be used in the space applications, with a particular emphasis on their MMOD impact resistance.

**Micrometeoroids and Orbital Debris**

The space environment imposes stringent and unique conditions that require careful and precise design of the spacecraft to ensure their longevity and functionality [21, 22, 23]. Hazards such as vacuum exposure, solar ultraviolet (UV) radiation, ionizing charged particle radiation, plasma interactions, surface charging and arcing, extreme temperature fluctuations, thermal cycling, MMOD impacts, and contamination can significantly degrade spacecraft materials and structure over time [21, 22, 23]. Consequently, the external materials of spacecraft must be engineered to withstand these harsh conditions, thereby preserving both structural integrity and operational performance [22, 24, 21, 2, 23].

In terms of MMOD impact protection, satellite design prioritizes safeguarding sensitive electronic components, communication systems, and power subsystems while keeping mass to a minimum—a critical requirement given the challenges associated with in-orbit repairs [16, 2, 25]. In contrast, manned spacecraft and space habitats must incorporate multifaceted protection systems that not only mitigate external impacts from micrometeoroids, orbital debris, and hyper-velocity collisions but also integrate robust thermal, structural, and environmental controls to ensure crew safety during all mission phases [26, 23]. While satellite shields typically rely on lightweight composite materials to achieve passive durability, human-rated vehicles require a more comprehensive design that balances mass constraints with active system monitoring, structural redundancy, and life support integration [27, 16].

**1. Need of Impact Protection**

The space environment is inherently hostile, requiring spacecraft to be meticulously designed for diverse objectives, including scientific exploration, commercial endeavors, and military applications, each necessitating varying operational lifespans [4, 19]. Spacecraft are strategically deployed into specific orbital regimes based on mission requirements [4, 28]. The three primary orbital categories used by humans—Geosynchronous Orbit (GSO), Medium Earth Orbit (MEO), and Low Earth Orbit (LEO)—each offer distinct functional advantages while introducing unique environmental challenges for spacecraft design, especially in terms of materials and systems [28].

Traditionally, telecommunications satellites were predominantly stationed in GSO due to its ability to provide consistent coverage over designated areas. However, with advancements in satellite technology, constellations such as Starlink and OneWeb have shifted to deploying satellites in LEO [28]. This change allows for reduced latency, lower launch costs, and accelerated orbital decay for non-functional satellites, addressing concerns about space debris accumulation. Nevertheless, LEO's lower altitude means satellites cover smaller areas, necessitating the deployment of larger constellations to achieve comprehensive global coverage [29, 30]. As of November 2024, there are over 5,500 Starlink and 630 OneWeb satellites in orbit, with projections indicating substantial growth in these constellations in the near future [29, 30, 31, 13].

This development underscores the evolving strategies and accelerated growth in satellite deployment to meet the demands of modern communications and ensure sustainability in the space environment. To illustrate the rapid growth, in late 2018, there were around 2,000 active Starlink satellites, and by July 2023, this number increased to 4,500, primarily in LEO [12]. In addition, Rwanda's Space Agency (RSA) has filed a request with the International Telecommunication Union (ITU) to place nearly 330,000 satellites in space as part of two constellations, Cinnamon 217 and Cinnamon 937, which were scheduled to launch by the end of 2023. However, they are not yet functional [13]. The reasons for the delay are not publicly available, but this trend clearly shows that the number of satellites in LEO is increasing rapidly.

Additionally, the average lifespan of these constellation satellites is five years, after which they are de-orbited and are supposed to burn up during re-entry into the atmosphere [32, 33]. This requires constellation operators to regularly replace them, which consequently necessitates frequent launches and intentional de-orbiting, resulting in a steady turnover in LEO and increasing the risk of abandoned satellites due to potential failures, such as improper de-orbiting or disposal [32, 33, 34].

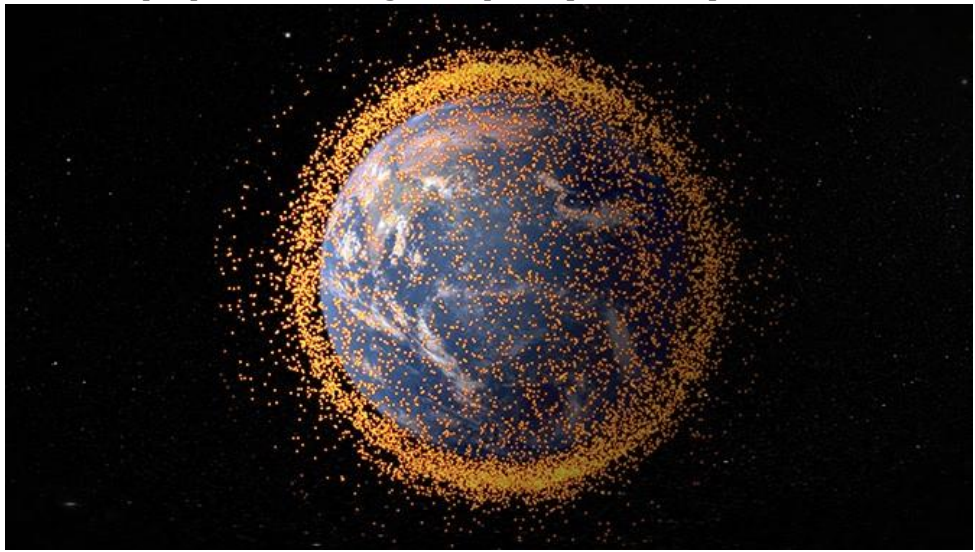

Figure 1: An artistic representation of the near-Earth orbital debris field, created using real data from the NASA Orbital Debris Program Office [35]

The image in Figure 1, sourced from [35], is an artistic representation of the near-Earth orbital debris field, created using real data from the NASA Orbital Debris Program Office. Created in April 2017, this visualization provides insight into the extent of debris in LEO. Given the significant increase in satellite launches and space activities in recent years, this depiction can be extrapolated to predict the growing concentration of orbital debris in LEO, emphasizing the escalating challenges in managing and mitigating space debris [35].

The plot of the projected growth of near-Earth debris under a "non-mitigation" scenario, taken from [36, 37], is illustrated in Figure 2. The data, averaged from 100 Monte

Carlo simulations with one sigma uncertainties, represents a worst-case scenario assuming no mitigation measures for current or future satellite launches [36, 37].

In this scenario, the debris population in LEO is expected to grow rapidly and nonlinearly over the next 200 years, highlighting the need for international and national mitigation measures that have been adopted in the last 15 years. In contrast, debris growth in MEO and GEO remains moderate, with only a few collisions between objects greater than 10 cm predicted over the same period [36, 37]. Current mitigation strategies, such as end-of-life maneuvers in GEO, effectively limit debris growth in these regions. Consequently, active debris removal (ADR) efforts should prioritize LEO, where the risk is significantly higher [36, 37].

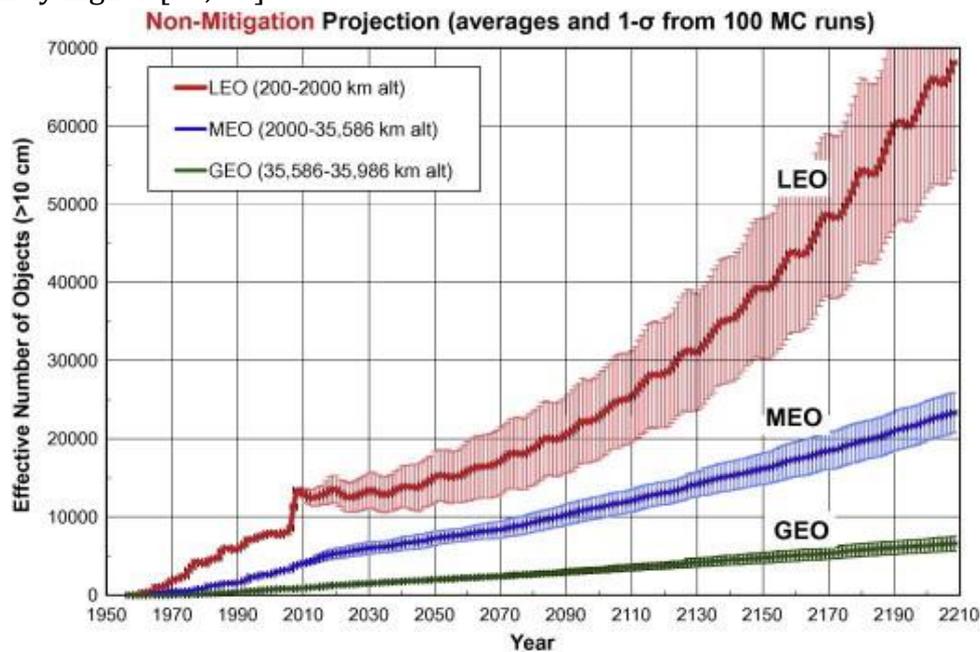

Figure 2: The projected growth of Orbital Debris in LEO, MEO, and GEO under a "non-mitigation" scenario[36]

A 2010 study predicted that large satellites in LEO would experience a mean lifetime reduction of approximately 13% over the next 50 years due to space debris, primarily caused by solar panel degradation from small, untrackable debris [34]. However, recent projections suggest that the deployment of over 5,000 new small satellites in high LEO orbits will amplify this impact. Collisions between tracked objects, small debris (down to 1 mm), and new satellites during replenishment activities could double the reduction in satellite operational lifetimes [34, 2]. In extreme cases, the degradation of solar panels alone could lead to a 60% reduction in the useful life [34].

Over time, these debris could accumulate, posing collision hazards to active satellites and contributing to space debris, which increases the overall risk of accidents and the challenges of managing the orbital environment [2]. Moreover, the deployment of these large constellations has raised significant concerns within the astronomical community, as these constellations affect the night sky views [2].

In [19], the rapid increase in space debris is highlighted, largely due to rising satellite launches and inadequate end-of-life disposal in LEO. The report mentions that for future moon or other deep space missions, it is crucial to keep low-Earth orbits safe to pass through for human space explorers, preserving cislunar space—the region between Earth and the Moon. Moreover, the same report states that the overcrowding of satellites in LEO escalates the risk of collisions, warning about the Kessler Syndrome if space debris are not taken seriously. In addition to this man-made space debris, there is always a chance of micrometeoroid impact to any spacecraft.

The Kessler Syndrome, also known as the Kessler Effect, is a theoretical scenario proposed by NASA scientist Donald J. Kessler in 1978 [6]. It is a potential scenario where the density of objects or debris in Earth's orbit becomes so high that collisions between them create a chain reaction of debris. Each collision generates more fragments, which then collide with other objects, producing even more debris. This self-sustaining process could ultimately make certain orbits unusable and pose severe risks to satellites, space stations, and future space missions [17, 6]. Kessler suggested that reaching this critical threshold could take 30 to 40 years. Some researchers argue that LEO, particularly within the altitude range of 900 to 1,000 kilometers, may have already reached a critical tipping point [6]. This concern is supported by Liou and Johnson in their study published in *Science* on January 20, 2006. According to them, "The current debris population in the LEO region has reached the point where the environment is unstable, and collisions will become the predominant mechanism for generating debris in the future" [36, 37]. This highlights an alarming shift, where Kessler Syndrome could dramatically increase debris levels, posing significant risks to space operations and sustainability.

A similar trend is highlighted in the *ESA Report* (2024), with Figure 3 from [19] supporting projections of catastrophic collision growth in Earth orbit. Despite improvements in mitigation strategies, a lack of compliance and limited remediation led to a net increase in space debris in 2023 [19]. Notably, 2023 saw the highest number of satellite launches on record [19]. If current activity patterns persist, catastrophic collisions are projected to rise sharply, as indicated by the steep increase in collisions through to 2225 [19]. By comparison, the scenario assuming no new launches after 2024 predicts significantly slower growth; however, this is an unlikely outcome given the ongoing rate of satellite launches [19].

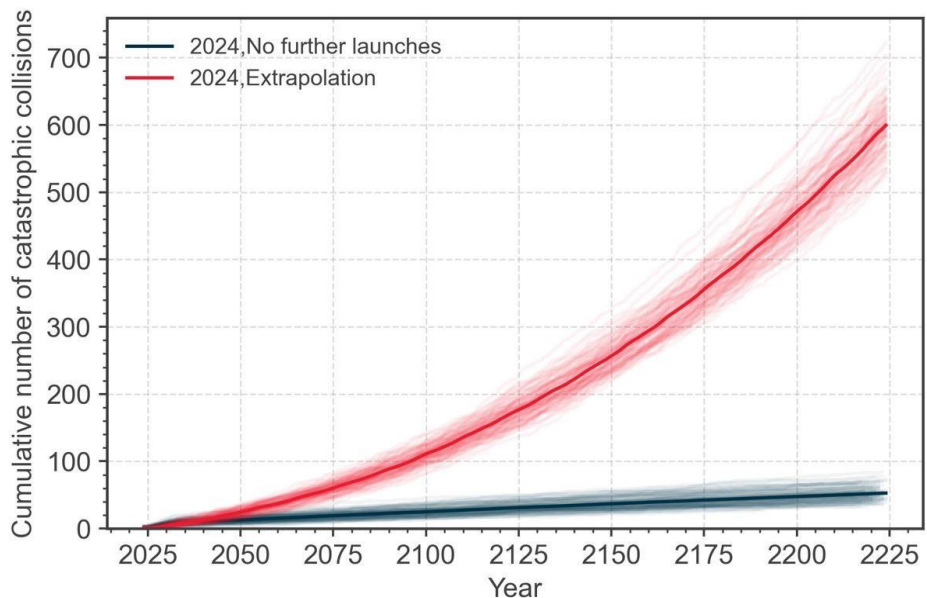

Figure 3: This illustration depicts the comparison of Cumulative number of catastrophic collisions if no further launches happens with the extrapolation of the current launches trend [19]

In addition to this man-made space debris, there is always a chance of micrometeoroid impact to any spacecraft [6]. Micrometeoroids are minuscule particles of rock or metal, often originating as fragments from larger celestial bodies. They typically range in size from 10 $\mu$m to 10 mm [38]. Particles exceeding this range are classified as meteoroids if they measure between 2 mm and 1 m, and as asteroids if they are larger than 1 m [38]. The following table shows a comparison between orbital debris and micrometeoroids, taken from [38].

Table 1: Comparison of Micrometeoroid and Orbital Debris[38]

| Comparison Items | Micrometeoroid | Orbital Debris |
| --- | --- | --- |
| Type | Naturally occurring | Human-made |
| Location | All locations in space or on the surface of vacuum bodies (Moon) | Currently limited to Earth orbit |
| Particle Size | 10 $\mu$m to 2 mm | Less than 1 cm to greater than 10 cm |
| Impact Speed | 11 km/s to 72 km/s | 1 km/s to 15 km/s |
| Altitude | Dominates below 270 km and above 4800 km | Dominates between 600 km and 1300 km |
| Trajectory | Travel to the Earth from interplanetary space | Orbit the Earth |

Although both micrometeoroids and orbital debris present significant hazards to spacecraft and space structures, their predominant altitudes are different. Orbital debris is generally confined to LEO, with the highest risk concentrated between altitudes of 600 and 1300 km [38, 10]. Beyond this range, orbital debris becomes less prevalent, and micrometeoroids emerge as the primary threat. Unlike orbital debris, which is restricted to specific orbital zones, micrometeoroids are present throughout space, predominantly below 270 km and above 4800 km [38, 10].

In Figure 4, taken from [10], the dominating areas of space debris and micrometeoroids are shown at the altitude ranges of debris and meteoroids domination, averaged over three orbital inclination angles: 90°, 0°, and 45°. There is considerable uncertainty in both meteoroid and debris flux measurements, and the precise locations of these regions are only approximate [10]. Therefore, the reader is advised to use these numbers as rough estimates.

The presence of orbital debris in LEO, coupled with the ever-increasing number of satellites, makes the protection of spacecraft critical. This is especially important as our day-to-day life on Earth depends largely on satellites. As shown in Table 1, the average speed of orbital debris impact is around 10 km/s, whereas for micrometeoroids, it can range from 11 km/s to 72 km/s. These high-speed impacts can not only damage spacecraft but can also be fatal for astronauts onboard.

Most traditional engineering designs aim to prevent plastic deformation during operation and reduce the risk of failures caused by fatigue. Consequently, the evaluation of mechanical properties for such applications typically involves static or cyclic testing, performed at strain rates of up to 1 per second, a range often referred to as the quasi-static regime [39]. However, certain applications demand materials to withstand plastic deformation at significantly higher strain rates, far beyond those encountered in quasi-static conditions. These scenarios are commonly classified as high strain rate (HSR) or dynamic impact applications. Safeguarding spacecraft against MMOD impacts is an example of an HSR application, where strain rates can reach up to $10^8$ per second [39].

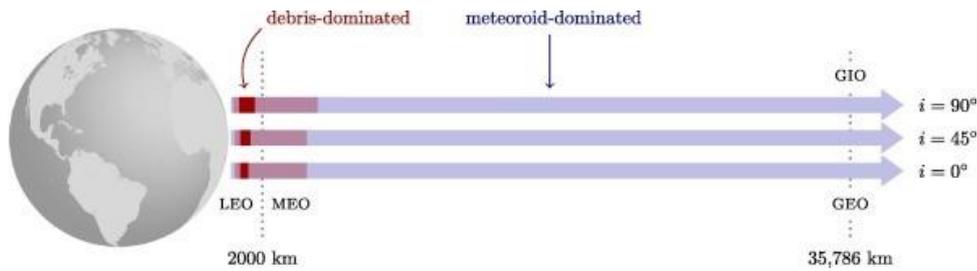

Figure 4: This illustration depicts the altitude ranges of the debris and meteoroids domination by averaging of three orbit inclination angles [10]

To provide perspective on the MMOD impact speeds in terms of strain rates, different regimes of strain rates are shown in Figure 5, taken from [39].

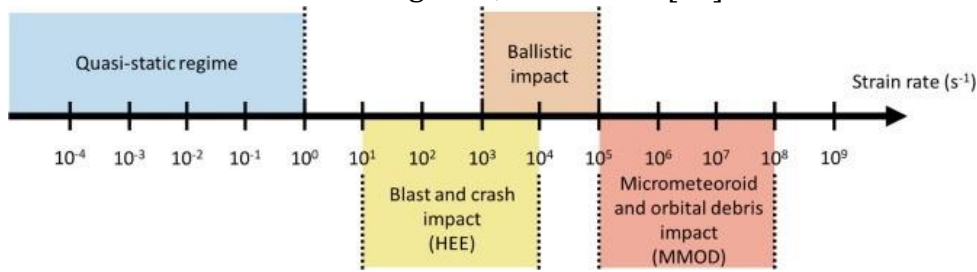

Figure 5: The different strain rates ranges from the quasi-static to dynamic applications. [39]

Additionally, space presents several other unique challenges for spacecraft materials, such as exposure to vacuum, intense ultraviolet radiation from the sun, ionizing radiation that results in material damage, electrostatic discharge effects, and thermal cycling (typically ranging from -175°C to 160°C) [22]. Atomic oxygen (AO) can erode spacecraft surfaces due to the high speeds (around 8 km/s) at which vehicles travel in LEO, resulting in significant kinetic impacts [22]. The degradation of spacecraft materials can cause substantial damage to the spacecraft and even pose a threat to the lives of astronauts in the case of manned spacecraft.

Figure 6 illustrates the damage sustained by Window No. 6 on NASA's Space Shuttle STS-126, caused by a meteoroid impact during its mission [40]. The window damage, the largest ever recorded on the shuttle, measured 13.43 mm by 8.97 mm, with a depth of 0.83 mm [40]. The meteoroid responsible for this damage was primarily composed of magnesium and silicon oxide, estimated to be approximately 0.2 mm in size, and traveling at an extraordinary velocity of 23 km/s [40]. This photograph was taken by the crew while the shuttle was in orbit [40].

On the left, Figure 6 depicts the damage to Window No. 6. On the right, it highlights the impact sustained by STS-118's Left-Hand Radiator No. 4, caused by orbital debris [40]. Based on impact tests and elemental analysis, the debris particle responsible was estimated to have a diameter of 1.6 mm and to have been traveling at a velocity of about 9 km/s [40].

This debris left a 6 mm diameter entry hole, penetrating completely through the radiator and into the payload bay door structure [40]. These images, taken from [40], demonstrate how spacecraft are prone to and affected by meteoroid and orbital debris impacts in space.

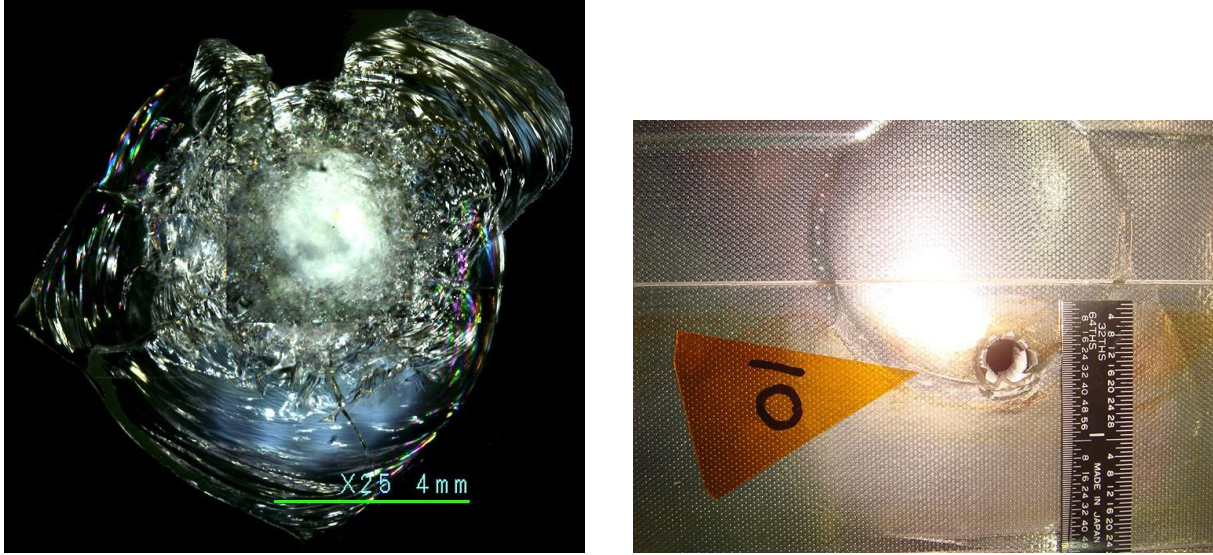

Figure 6: The left image shows the largest recorded window damage on STS-126, while the right highlights the radiator damage on STS-118. [40]

*1.1. Key Approaches to Micrometeoroid and Orbital Debris Mitigation*

To mitigate the challenges posed by the impacts of MMOD, various strategies are employed, as discussed in the preceding section. These approaches are broadly classified into two categories of protection:

- Active methods

- Passive methods

*1.1.1. Active Methods of MMOD Protection*

Active methods employ dynamic measures to reduce the risk of MMOD impacts. This is considered one of the best approaches to preserving the space environment, as no debris is generated in this process [38]. These strategies include:

1. **Evasive Maneuver:** Evasive maneuvers are vital strategies executed by spacecraft to avoid potential collisions with space debris, safeguarding mission objectives and the spacecraft's structural integrity [41, 38, 19]. These maneuvers are planned based on predictive models, such as NASA's Conjunction Assessment Risk Analysis (CARA) [41] and ESA's MASTER (Meteoroid and Space Debris Terrestrial Environment Reference) [42], which monitor debris movements and provide predictions of conjunctions. The process begins with continuous debris tracking using radar and optical systems, generating data on the orbits of potential collision threats [41]. Once a conjunction is identified, the risk is analyzed based on parameters such as relative velocity, proximity, and size of the debris [42]. If the collision risk surpasses predefined thresholds, an evasive maneuver is calculated, typically performed using thrusters to alter the spacecraft's position or velocity [41, 38, 19]. For spacecraft like the ISS, these maneuvers are routine due to its location in a heavily trafficked orbital region [41]. The frequency of such operations has increased, highlighting the necessity for advancements in debris monitoring systems [42, 19].
2. **Deorbiting Decommissioned Spacecraft:** Deorbiting decommissioned spacecraft is a proactive measure to mitigate the risks posed by orbital debris [19]. By actively removing non-functional satellites and other debris from orbit, the density of hazardous objects is reduced, thereby minimizing the likelihood of collisions with operational spacecraft [43, 19]. This approach is supported by initiatives such as ESA's Deorbit Kit, which enables satellites to perform propulsive decommissioning maneuvers at the end of their mission or after a failure [43]. Additionally, NASA's Orbital Debris Mitigation Standard Practices (ODMSP) emphasize the importance of post-mission disposal, including atmospheric reentry or other methods, to ensure long-term sustainability in space [44].

*1.1.2. Passive Methods of MMOD Protection*

Passive methods for mitigating the effects of MMOD impacts emphasize the design of spacecraft systems that are inherently resilient [16, 45]. These methods aim to reduce the risk of catastrophic damage by incorporating robust materials, structural reinforcements, and innovative shielding technologies [16, 45, 46]. Similarly, ESA's guidelines on MMOD protection highlight the importance of designing spacecraft with durable structures to withstand impacts [42]. These approaches are critical for ensuring the long-term sustainability of space missions, especially in the increasingly crowded orbital environment.

1. **Enhanced Shields:** In this strategy, advanced shielding technologies, such as Whipple shields, multi-layer materials, and lightweight composite shields, play a critical role in safeguarding spacecraft against the impacts of MMOD [16, 45, 46]. Whipple shields, for instance, are designed to dissipate the energy of high-velocity particles by employing a sacrificial outer layer called a bumper that reduces the impact force before it reaches the main structure [16, 45, 46]. Multi-layer materials add further protection by using multiple barriers of varying thickness and properties, effectively dispersing impact energy over a larger area [16, 45, 46]. Lightweight composite shields enhance this design by combining strength and flexibility while minimizing additional weight, which is essential for maintaining spacecraft efficiency [16, 45, 46].

   NASA's HVIT group has conducted extensive research into MMOD protection, focusing on passive shielding techniques such as Whipple shields and multi-layer materials [45]. Their findings highlight the effectiveness of these technologies in mitigating damage from hypervelocity impacts, which are increasingly common due to the growing density of debris in Earth's orbital environment [45]. These advancements not only ensure the safety of individual missions but also contribute to the long-term sustainability of space operations by reducing the risk of catastrophic damage from MMOD collisions [16, 45, 46].

2. **Durable Structures:** The design of spacecraft with robust and resilient structures is critical for ensuring protection against MMOD impacts [16, 47, 48, 49]. By utilizing advanced materials like high-strength alloys and composite materials, spacecraft can withstand HVIs and continue functioning as intended throughout their lifecycle [47, 49]. For instance, NASA's studies on spacecraft durability emphasize the importance of resilient structural integrity in minimizing mission-critical failures [48]. Such durable designs are particularly significant for long-term missions in heavily trafficked orbital regions, where MMOD risks are prevalent [16, 47, 49].

3. **International Guidelines and Policies:** International cooperation and adherence to established guidelines are essential for mitigating the risks posed by orbital debris [50, 51, 49, 52]. Guidelines such as the United Nations' Space Debris Mitigation Guidelines and policies from space agencies like ESA and NASA set standards for responsible orbital behavior and the sustainable use of space [50, 51]. These standards include strategies for debris mitigation, collision avoidance, and postmission disposal. Following these best practices not only reduces the risks of debris collisions but also promotes a sustainable orbital environment for future generations [52, 53].

   According to *ESA Report* (2024), in 2023, the total number of reentries decreased due to the prior peak of debris from a 2021 anti-satellite missile test [54]. However, the number of satellite (payload) reentries rose sharply, driven by increased compliance with space debris mitigation guidelines [19]. Both payload and rocket body reentries continue to grow annually, influenced by improved end-of-life disposal practices and heightened solar activity, which accelerates atmospheric drag and reentry. Notably, ~ 90% of rocket bodies in LEO complied with disposal standards, with over half reentering in a controlled manner, as shown in figure 2.18 [19].

*1.1.3. Examples of Current Passive Methods in Spacecraft Structures*

The most common MMOD (Micrometeoroid and Orbital Debris) shield used in various configurations is the Whipple shield [16, 55, 46, 45]. This shield utilizes a multi-layered design where the outer bumper layer breaks incoming debris into smaller fragments, while subsequent layers disperse the kinetic energy of these fragments [16, 55, 45].

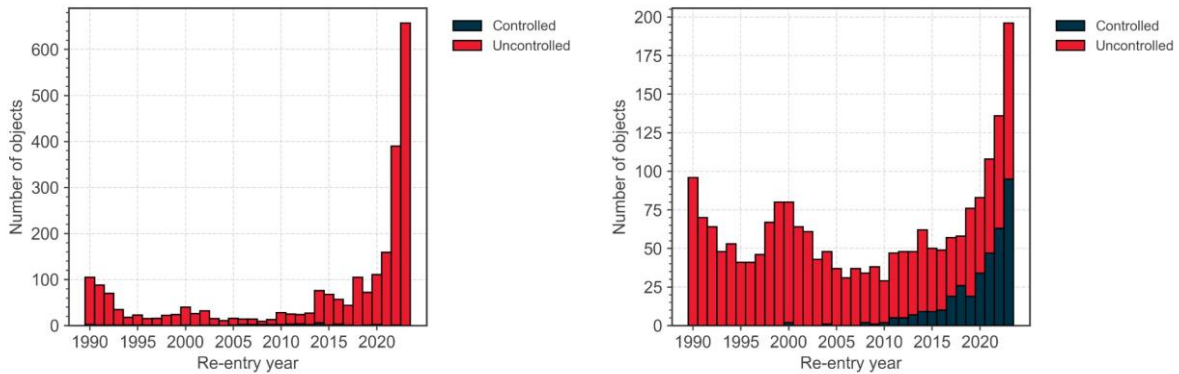

Figure 7: Annual reentries of payloads (displayed on the left) and rocket bodies (displayed on the right) into Earth's atmosphere, categorized by controlled and uncontrolled methods.[19]

This concept has been effectively implemented on spacecraft such as the ISS, providing enhanced protection in critical areas [16, 55, 45]. Whipple shields were specifically designed for use in sections of the ISS where meteoroid impacts are anticipated. They are not applied throughout the entire station due to their considerable weight, which poses design and operational constraints [16]. Several types of shields have undergone testing, and efforts are actively underway to improve their design and functionality. Below are some examples of these shields:

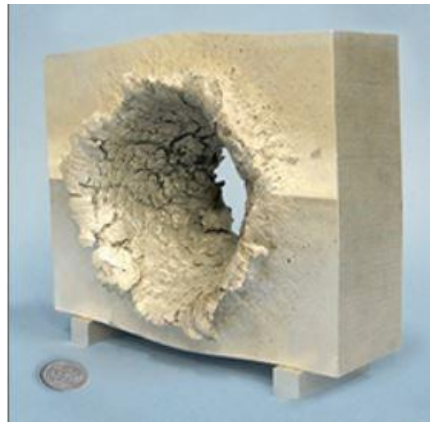

Figure 8: Monolithic Shield[45]

**Monolithic Shield:** The Monolithic Shield represents the most basic form of MMOD protection. It consists of a single slab of aluminum capable of absorbing the full force of an impact. While simple in design, it provides robust defense against high-velocity debris [56]. This shielding method is primarily used as a reference for comparing advanced shielding solutions with equivalent mass and is often considered the "default" protection option for spacecraft structures [56]. Due to its simplicity, the monolithic shield serves as a benchmark to evaluate more innovative solutions.

**Whipple Shield:** The Whipple Shield is the first spacecraft shield ever implemented. It was introduced by Fred Whipple in the 1940s and is still in use today. The basic design consists of placing a sacrificial bumper, usually made of aluminum, in front of the spacecraft, allowing it to absorb the initial impact [56, 57]. Upon impact, the Whipple bumper shocks the projectile and generates a debris cloud composed of smaller, less lethal fragments from both the bumper and the projectile. This debris cloud spreads over a larger area, reducing the energy concentration upon reaching the spacecraft's rear wall [56, 57].

**Stuffed Whipple Shield:** The Stuffed Whipple Shield is an enhanced version of the basic Whipple design. NASA and ESA introduced the Stuffed Whipple Shield (SWS) for spacecraft modules like the U.S. Laboratory and Columbus [57]. It incorporates layers of advanced fabrics such as Nextel and Kevlar between the bumper and the rear wall. These additional layers help to further disrupt and pulverize the incoming debris cloud, ensuring that any fragments reaching the rear wall are sufficiently weakened to pose minimal threat [56, 57].

**Multi-Shock Shield:** The Multi-Shock Shield is a commonly used shielding design that utilizes multiple staggered layers of Nextel positioned at specific standoff distances. Each layer incrementally shocks and disperses the projectile and its resulting debris cloud. This progressive degradation continues until the remaining fragments are too harmless to penetrate the rear wall [38, 56, 57].

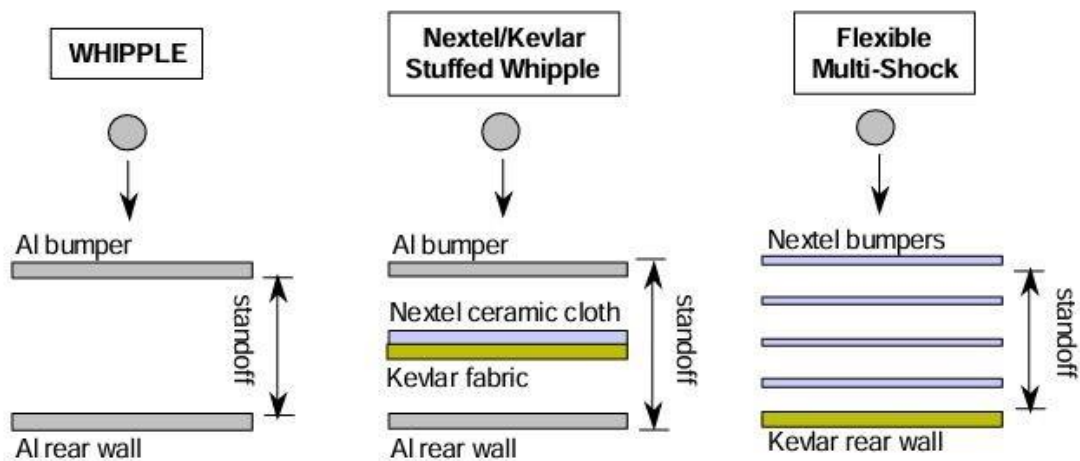

Figure 9: MMOD Shield types[45]

**Mesh Double Bumper Shield:** The Mesh Double Bumper Shield features a doublelayered aluminum mesh bumper placed ahead of an aluminum rear wall. This configuration offers an effective balance of shock absorption and fragmentation dispersion, making it a lightweight and efficient option for certain spacecraft applications [38, 56, 57].

**Multifunctional Structure:** In addition to these direct shields, another strategy being employed for spacecraft protection is known as the Multifunctional Structure (MFS). An MFS integrates multiple subsystems with the load-bearing structure to reduce overall system mass and volume, improving space system efficiency and lowering mission costs [58]. This integration includes components like batteries, thermal control systems, and electronic subsystems. MFS technology is particularly useful for small satellites and interplanetary missions, where weight and volume constraints are critical. The reduction in structural mass allows for more payload, enhancing scientific outcomes. The structure also provides additional functionalities such as radiation protection, thermal control, structural health monitoring, and protection against micrometeoroids and orbital debris [58]. These features make MFS highly suitable for space exploration and crew launch vehicles, as they enhance survivability and lifespan by minimizing impacts.

**Foam-Based Shields:** These shields utilize lightweight foam-like materials to absorb and dissipate impact energy. Current advancements are exploring the integration of aerogels, which provide enhanced thermal insulation and impact resistance. Notable shielding components include Honeycomb Core Sandwich Panels (HCSP) and Foam Core Sandwich Panels (FCSP), often incorporated into spacecraft walls [46].

HCSPs consist of metal honeycomb cores bonded to metal or woven face sheets, offering exceptional structural integrity. However, these panels exhibit reduced shielding efficiency due to the *channeling effect*[1], where debris remains intact while passing through the honeycomb core [46]. In comparison, FCSPs employ metal foam cores that mitigate the channeling effect, enabling more uniform dissipation of impact energy across all directions. Shielding performance is further improved by employing fluid- or gas-filled honeycomb cores situated between bumper plates and rear walls, which redistribute impact energy and vaporize projectiles [46].

Leading MMOD shields, such as the Whipple shield, rely on thin plates separated by void spaces to intercept and fragment incoming debris. Adding intermediate layers, such as fabric or additional bumper plates, has demonstrated significant improvements in shielding effectiveness. Nevertheless, these modifications often result in increased non-ballistic mass for installation components like fasteners and supports, which can constitute up to 35% of the total shield weight [46].

Furthermore, [46] investigated metallic foam sandwich structures for MMOD protection. Tests conducted across varying impact angles revealed superior shielding performance compared to traditional honeycomb designs, making metallic foam structures a promising alternative for spacecraft shielding applications.

**Hybrid Shield Designs:** Hybrid shield designs aim to merge metal and polymer layers to achieve an optimal balance between durability and weight while enhancing resistance against HVIs. Currently, the SpaceX Dragon commercial cargo vehicle employs cost-effective fiberglass fabric instead of more expensive ceramic cloth for protection against MMOD particles [60]. This innovative approach demonstrates potential for broader applications, including future inflatable modules. Furthermore, Orbital Sciences is evaluating this technology for integration into the Cygnus commercial cargo vehicle. Other spacecraft under development may also adopt this cost-effective and efficient shielding solution for MMOD protection [60].

Another type of shield design involves self-sealing mechanisms that are lightweight and capable of effectively sealing holes with a diameter of 6.4 mm in the pressure shell under a pressure differential of 1 atmosphere, mimicking conditions found in the ISS crew modules. One approach involves the use of ionomer plastics combined with elastomeric materials, which has been successfully validated through HVI testing to seal penetrated pressure shells under a 1 atmosphere pressure differential [61]. Advancements have been achieved in MMOD shielding, surpassing the performance of conventional ISS stuffed-

---

[1] Channeling effect: Honeycomb cores, with their hexagonal cell structure, can restrict the radial expansion of debris clouds formed from HVI and guide them along the axial path [59].

Whipple shields and integrating enhanced radiation protection. Furthermore, self-sealing MMOD shield designs have been successfully validated through HVI testing [61].

## 2. Structure and Materials

Space agencies around the globe are actively developing new strategies to safeguard their assets in space against Hyper-velocity impacts (HVI) caused by MMOD. These efforts not only aim to protect critical space assets but also ensure mission success and the safety of onboard crew members [45]. These MMOD threats, though distinct in nature, present significant risks to spacecraft structures and crew safety. To mitigate these risks, the design of spacecraft capable of withstanding high-speed MMOD impacts has become a critical focus.

*2.1. Materials Selection Standards*

In the previous section, various structural designs of MMOD shields were analyzed, highlighting their effectiveness in providing spacecraft protection. This sub-section focuses on the selection standards for the material compositions used in spacecraft. These materials are selected based on their ability to meet the operational requirements of spacecraft while minimizing mass and ensuring reliability. The emphasis is always on optimizing both material selection and configuration to enhance protection against HVIs while minimizing overall mass. Some of the key criteria that are checked when selecting a spacecraft material are as follows:

**Flammability, Offgassing, and Compatibility:** Materials for habitable and non-habitable spacecraft zones must meet stringent flammability, odor, offgassing, and compatibility thresholds (total mass loss 1.0 % 1.0%, condensable volatiles 0.1 0.1%) under ASTM E-595, as codified in NASA-STD-6001 [62]. This document outlines that the worst-case anticipated use environment should be considered when evaluating material suitability. Materials meeting acceptance criteria are considered suitable for design consideration. Additionally, it is noted that materials already meeting test criteria should be selected when possible [62].

**Electrical Charging Considerations:** The selection process also incorporates surface and internal charging performance, surface resistivity, dielectric breakdown, and low secondary electron yield following the design guidelines in [63] to mitigate spacecraft charging anomalies in diverse plasma environments [63]. These properties are critical in ensuring that the materials can handle prolonged exposure to plasma-rich environments without accumulating dangerous levels of charge. Improper material selection can lead to phenomena such as arcing, which may damage sensitive spacecraft components or result in mission failure [63].

Guidelines in [63] outline best practices for reducing space charging effects by recommending materials and configurations that dissipate charge efficiently and resist dielectric breakdown under high-voltage conditions. Incorporating these considerations into material selection enhances the spacecraft's overall reliability, ensuring safe and uninterrupted operation in various orbital and plasma environments. These measures are vital for satellites and spacecraft operating in LEO, GEO, or interplanetary missions, where plasma interactions are particularly pronounced [63].

**ESA Preliminary Guidance:** ESA's PSS-01-701 offers data-driven thresholds for mechanical, thermal, and chemical properties during early material trades, orienting European programs toward candidate materials that satisfy cross-disciplinary requirements

[64].

**Structural and Thermal Control Materials:**

**Metallic Alloys and Composites** Over five decades of flight heritage underscore aluminum (2xxx, 5xxx, 7xxx series, Al–Li), titanium, and stainless steels for primary structure, augmented by graphite/epoxy and carbon-fiber polymer-matrix composites to optimize strength-to-weight ratios and stiffness lessons compiled in Finckenor's chapter on spacecraft materials [24].

**Multilayer Insulation:** Thermal management employs multilayer insulation assemblies per NASA/TP-1999-209263, which catalogues foil, spacer, and support materials for passive temperature control on platforms from Spacelab to ISS [65].

**Pressure Shell and Composites** SpaceX's Dragon employs an aluminum-alloy pressure vessel with carbon-fiber composite panels to reduce mass while maintaining structural integrity, consistent with NASA material-screening practices [24].

**Thermal Protection Systems:** Dragon's heat shield leverages NASA's PICA technology refined into the PICA-X variant for manufacturability and reusability, first flight-tested in COTS Demo 1 (2010) and validated via post-flight analysis of returned samples [66] [67]. SpaceX additionally developed a proprietary ablative insulation (SPAM) for the backshell, integrating vertical in-house fabrication of raw PICA blocks sourced from FMI. In addition to all these guidelines, various material compositions of MMOD shields are analyzed, highlighting their effectiveness in providing spacecraft protection. NASA's Handbook for Designing MMOD Protection (SP-20090010053) prescribes risk-based design using ORDEM environment models and ballistic-limit equations to size Whipple bumpers, with stuffed Whipple variants (Kevlar/Nextel interlayers) offering enhanced fragmentation and energy absorption [24].

*2.2. Materials Used in Shields*

This section focuses on the material composition of these shields, emphasizing the importance of optimizing both material selection and configuration to enhance protection against HVIs while minimizing overall mass. These materials are selected based on their ability to meet the operational requirements of spacecraft while minimizing mass and ensuring reliability.

Aluminum alloys such as Al 6061-T6 are commonly employed for bumper plates because of their ability to fragment impacting projectiles and facilitate debris expansion [16, 37, 38]. NASA has explored the use of titanium alloys and impedance-graded materials, which, in some cases, outperform homogeneous aluminum in resisting HVIs [16]. Additionally, the deployment of multi-layer insulation (MLI) wraps on aluminum bumpers, using materials such as Teflon-coated fiberglass, aluminized Mylar, and Kapton, has shown robust shielding performance [25].

Further investigations have examined metal foams (including those based on aluminum, copper, nickel, steel, and titanium) with varying porosities, as well as fiber-reinforced polymers (FRP) to enhance bumper behavior under impact [25, 46]. Rear walls are typically fabricated from aluminum alloys (e.g., Al 6061 and Al 2024) due to their favorable strength-to-weight ratios, while titanium alloys are used where higher tensile strength is necessary [16]. Hybrid composite rear walls incorporating materials such as carbon fiber-reinforced polymer (CFRP) and polycarbonate have also been tested for improved energy absorption.

For intermediate stuffing layers, ceramic fabrics (e.g., Nextel) and aramid fabrics

(e.g., Kevlar) are popular choices because they combine high impact resistance with excellent tensile strength [25, 46]. Other materials, including open-cell metal foams, aluminum honeycomb, pinewood, and glass-fiber-reinforced aluminum laminate (Glare), have demonstrated weight-efficient shielding capabilities [38]. Flexible materials, such as polyimide foam and polymer batting, are used to fill the space between bumpers and rear walls, providing expansion zones that enhance shield performance by facilitating energy dissipation from projectile fragments [38].

Through adherence to rigorous aerospace standards and incorporation of both heritage and novel materials, modern spacecraft, from ISS modules to commercial vehicles like Dragon, achieve optimized structural performance, thermal control, and MMOD protection. Continued advancements in composites, ablators, and multifunctional shields promise further mass savings and mission safety enhancements.

In addition to the application of shields, there are other methods that are also very effective and play a crucial role in spacecraft protection against MMOD. MMOD analysis is conducted using tools such as BUMPER, which help determine the probability of spacecraft damage or failure due to meteoroid and debris impacts, and enhance spacecraft survivability [16]. This task requires a comprehensive understanding of spacecraft systems, including:

- Operations and failure modes,

- Subsystem design specifics,

- Materials of construction,

- Operational parameters.

This is because MMOD impact protection is essentially a complex system function, and spacecraft are often very intricate systems. Thus, it is crucial for the MMOD analyst to understand the system requirements as mentioned above and prioritize the selection of the most capable protection solutions that fit within spacecraft and program constraints, including weight, cost, schedule, and risk.

*2.3. Light-weight Shields*

In the MMOD environment, spacecraft are highly susceptible to damage from highvelocity impacts, as previously discussed. These impacts can degrade spacecraft performance, shorten mission durations, or even result in catastrophic failure. To mitigate such risks, NASA establishes rigorous MMOD protection requirements tailored to individual spacecraft [16]. These requirements encompass both design and operational considerations, both of which are critical to ensuring survivability in the hostile MMOD environment [16].

Several factors influence the design and implementation of MMOD shielding systems, including mission duration, spacecraft dimensions, orbital altitude and attitude, shield type, and the assessed risk level [16]. Additionally, spacecraft weight and volume constraints often limit the extent to which dedicated MMOD shielding can be deployed [16].

The graph in Figure 10 illustrates the relationship between the level of MMOD protection required and the corresponding shield weight. As the level of required protection increases, the necessary shield mass also increases, often at an accelerating rate. This trend highlights a fundamental design trade-off: maximizing spacecraft protection while minimizing added weight, which is a critical consideration in both spacecraft design and mission planning.

Most smaller satellites are not equipped with MMOD shields due to the relatively low likelihood of MMOD collisions, which is attributed to their smaller size and shorter mission durations [16, 46]. Furthermore, incorporating MMOD shielding would significantly increase satellite mass, thereby raising launch costs and reducing cost-efficiency for such missions. As an alternative, the primary structure of the satellite is designed to provide the necessary level of shielding [16, 46]. Typically, these primary structures are composed of honeycomb core sandwich panels (HCSPs), which offer excellent specific strength and stiffness. However, HCSPs exhibit limited protective capabilities against MMOD impacts, necessitating the implementation of additional protective measures [16, 46].

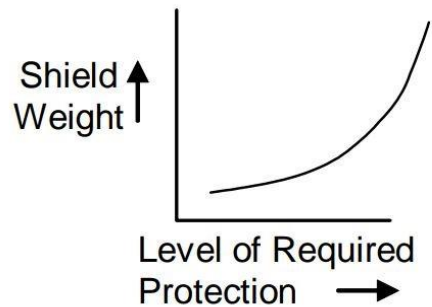

Figure 10: Relation between Shield weight and level of protection[16]

In spacecraft design, minimizing the weight of MMOD shielding is critical to optimizing overall performance and cost-efficiency, while still maintaining adequate protection for both the spacecraft and crew. To achieve this balance, a strategic trade-off is employed in the design of shielding systems based on location-specific risk exposure [16]. This approach involves constructing spacecraft with variable shielding thicknesses: regions with a higher probability of MMOD impact receive enhanced protection, while areas with lower impact risks are outfitted with more moderate shielding [16].

The distribution of shielding is determined through a detailed analysis of the spacecraft's orbital trajectory and its orientation relative to the velocity vector, as these factors strongly influence the likelihood of MMOD collisions [16]. As shown in Figure 11, the BUMPER geometry model of the International Space Station (ISS) from 2006 illustrates this concept. Each color represents a different shield type, excluding the solar arrays. The finite element model (FEM) of the ISS geometry includes approximately 150,000 elements, with an average element size of 20 cm × 20 cm [16].

This targeted allocation of shielding enables a balanced trade-off between safety and weight optimization, enhancing both protection and mission efficiency. By strategically distributing shielding based on localized risk, designers ensure optimal safety without incurring unnecessary mass penalties [16].

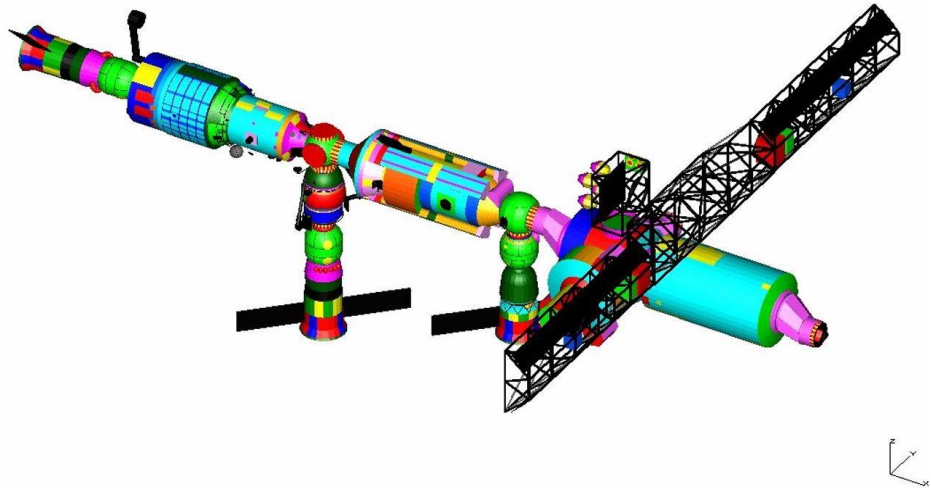

Figure 11: The BUMPER Geometry Model of the ISS, 2006. Each color denotes a different shield types [16].

An example of risk mitigation through spacecraft orientation is the adjustment of the Shuttle–ISS mated flight alignment during the ISS docked phase, as illustrated in Figure 12. By transitioning from the +XVV to the −XVV [2] orientation, the Shuttle reduced its exposure to MMOD.

This adjustment reduced MMOD risks to vulnerable areas, such as the wing leading edge (WLE) and nose cap (NC), by a factor of 5 [16]. Prior to STS-114, the Shuttle faced the ram velocity direction[3], exposing it to higher MMOD flux [16]. The orientation shift redirected sensitive surfaces away from the particle flux, enhancing both crew safety and mission success. It is worth highlighting that STS-114 was NASA's first "Return to Flight" mission after the space shuttle, *Columbia* tragedy in 2003 [68]. This mission introduced the innovative rendezvous pitch maneuver for damage assessment, setting new safety standards for future flights [68].

As humanity endeavors to establish colonies on the Moon and Mars, the frequency and duration of space missions are projected to rise significantly, underscoring the critical necessity of developing lightweight MMOD shields to ensure mission success and efficiency [38].

## 3. Additive Manufacturing

Additive Manufacturing (AM), also known as 3D printing, is a collection of processes that create components layer by layer from digital models [69, 70, 71, 72]. This approach enables the production of complex or customized parts directly from design files, eliminating the need for expensive tooling or molds. This paradigm shift reduces part counts, minimizes assembly requirements, and supports on-demand production, lowering inventory costs and lead times for industries such as aerospace, healthcare, energy, and automotive [69]. Notable applications include integrated aerospace fuel nozzles, patient-specific medical implants, and high-efficiency burner tips, all leveraging AM's capacity to bypass the design constraints of traditional methods [69].

---

[2] • **+XVV:** Points in the direction of the velocity vector, representing the direction the ISS is moving towards.

• **-XVV:** Opposite to +XVV, pointing in the direction from which the ISS is coming.

[3] • **Ram Velocity direction:** Velocity direction of ISS motion.

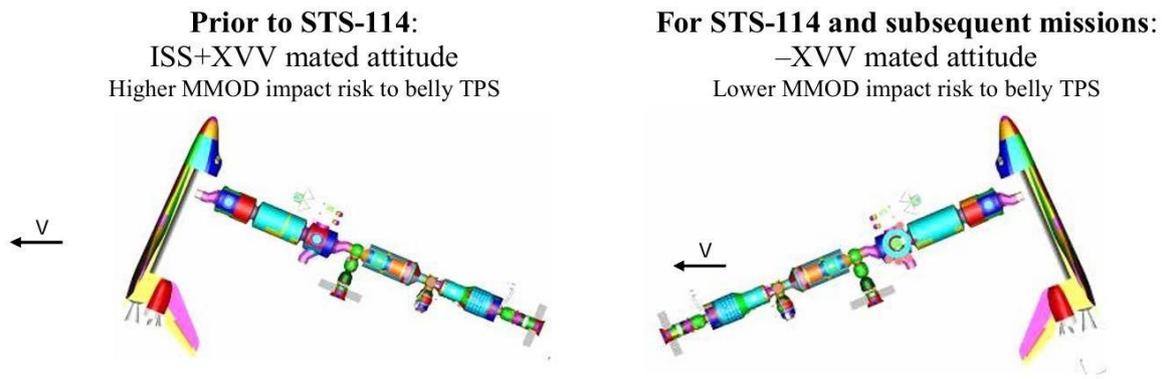

Figure 12: Shuttle docked in varied orientations with ISS before and during STS-114 mission to assess in-orbit the damage to the shuttle during launch, and minimize MMOD risks [16]
.

Emerging from rapid prototyping roots, AM has evolved into a viable production method for structural metallic components, driven by advancements in laser systems, computational power, and metal powder technologies [69]. While commercial AM systems now achieve critical industrial acceptance, certification remains largely part-specific, necessitating deeper insights into feedstock-process-property relationships to ensure defectfree, reliable parts [69].

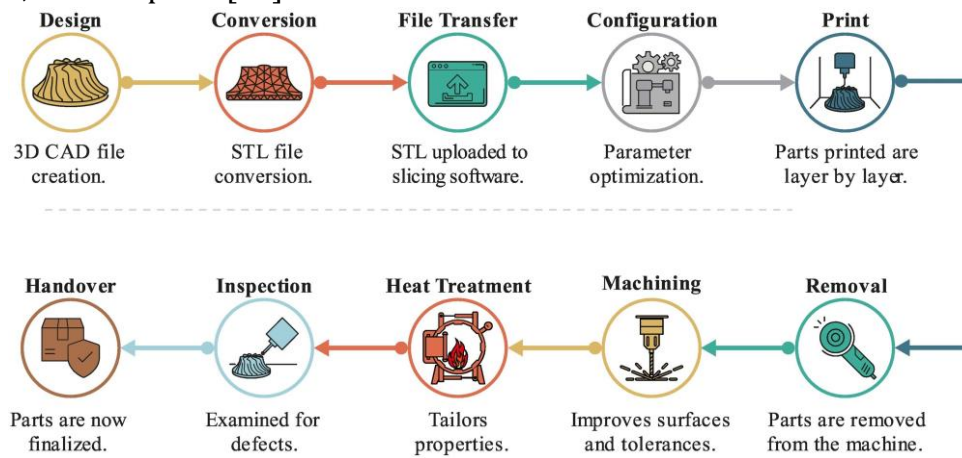

Figure 13: Additive Manufacturing Process[73]

According to ASTM International's F42 committee, AM encompasses seven primary process categories: material extrusion, vat photo-polymerization, powder bed fusion, binder jetting, direct energy deposition, sheet lamination, and material jetting [74]. Each method offers distinct benefits and limitations depending on the material system, resolution, and functional requirements [72, 69, 70, 71, 74].

The fabrication of metal structures using Metal Additive Manufacturing (MAM) involves multiple stages. The specific workflow for a product can vary depending on its design and intended application [73]. For instance, a prototype component might only require printing to assess its feasibility before scaling up production. Conversely, parts intended to substitute high-value, structurally critical components typically undergo extensive post-processing treatments and rigorous quality inspections before deployment. Nonetheless, the overarching processes can be consolidated into a single flow diagram, as depicted in Figure 13 [73].

### 3.1. Types of Additive Manufacturing

**Vat Photo-polymerization**: This process uses a vat of liquid photopolymer resin, cured by a light source (usually UV), to form solid layers [71, 72]. Vat photo-polymerization encompasses three techniques—stereolithography (SLA), digital light processing (DLP), and continuous DLP (cDLP)—which use light sources to cure photopolymer resin within a reservoir [71, 72]. SLA employs a laser to trace resin layers, with top-down or bottom-up printer orientations influencing the build process. DLP and cDLP utilize digital projectors to cure entire layers at once, with cDLP enabling faster production through continuous z-axis movement [71].

Vat photo-polymerization offers exceptional detail, surface finish, and versatile resin options for applications ranging from jewelry and dental models to intricate prototypes [71]. It is particularly beneficial for healthcare applications, including anatomical models, drug delivery systems, surgical guides, and biocompatible implants, leveraging materials that meet stringent safety standards [71].

VP's advantages include fast production speeds, reduced support requirements, and high precision, though limitations such as restricted build volumes, material costs, brittleness, and potential environmental concerns remain challenges [71]. With advancements in resin properties and processing techniques, VP continues to revolutionize industries, promoting innovation in medical treatments and personalized manufacturing [71].

**Material Jetting**: Similar to inkjet printing, this method jets droplets of material onto a build platform, which are then cured or solidified. Material Jetting (MJ) utilizes a print head to deposit heated photopolymer beads onto a build platform, forming layers solidified by UV light to create intricate shapes with high resolution [71, 72]. The system supports multi-material printing, offering flexibility for simultaneous deposition of diverse materials, including soluble supports. This technology enables layer thicknesses as fine as 16 microns, ensuring exceptional detailing and smooth surfaces, particularly in applications like vibrant prototypes, architectural models, and biocompatible medical devices [71].

MJ machines allow for rapid prototyping and small-batch production with minimal support structures that are easily removed during post-processing. However, MJ faces limitations such as high initial and operational costs, restricted build volumes, material restrictions, and material waste due to support removal [71]. The technology requires skilled operation and regular maintenance, adding complexity. Prominent brands like Stratasys and 3D Systems dominate MJ systems, applying it in industries from healthcare to manufacturing [71]. Emerging trends include innovations such as Nano Functional Technology and Drop-on-Demand systems, which further expand MJ's applicability in producing detailed wax models for jewelry casting and high-speed tooling for industries

[71]. MJ remains transformative across various sectors, advancing precision and versatility in 3D printing applications [71].

**Binder Jetting**: Binder Jetting (BJ), developed at MIT in the 1990s, employs a nozzle to deposit liquid binder onto a powder bed, forming cross-sectional layers that solidify through curing [71]. This layer-by-layer approach enables the creation of complex geometries with high precision, supported by controlled binder droplet sizes under 100 microns [71, 72]. The resulting "green body" requires post-processing, including curing, infiltration, and sintering, to enhance structural integrity and surface finish.

BJ excels in cost-effectiveness, scalability, and versatility, supporting materials such as metals, ceramics, and polymers [71]. Unique features include minimal heat-induced defects and the ability to produce multicolor parts, broadening applications in prototyping, functional components, and industries like pharmaceuticals, dental medicine, and aerospace. Challenges include lower density, labor-intensive post-processing, deformation risks, binder residue, and limited mechanical properties of green bodies [71]. Recent advancements aim to reduce porosity, with BJ now being employed for applications such as producing customized oral medications like Spritam, improving accessibility and affordability [71]. BJ continues to revolutionize manufacturing by pushing boundaries across medical and industrial landscapes [71].

**Material Extrusion**: Also known as Fused Deposition Modeling (FDM), this is a unique production method where print orientation and raster design angle must be chosen carefully to reduce strain on objects [71, 72]. PLA is commonly used, and printing involves extruding two materials—one for the part and another for support—through heated extrusion heads at 190–230°C [71]. Thermoplastics are printed in layers, fusing upon cooling, using a bottom-up approach controlled across x, y, and z axes. Finished parts are removed by twisting the tray, and supports are detached manually [71].

FDM is advantageous for its affordability, reproducibility, speed, and low maintenance, though it faces limitations like material waste, production delays due to supports, and difficulty handling complex geometries [71]. Its applications span industries, enhancing healthcare with anatomical models and medical devices, aiding automotive manufacturing with lightweight components, and empowering hobbyists and educators for diverse projects [71, 72]. Despite its limitations, FDM remains transformative, offering efficiency, innovation, and broad accessibility across sectors [71, 72].

1. **Powder Bed Fusion (PBF)**: This AM category includes processes such as Selective Laser Sintering (SLS) and Selective Laser Melting (SLM), in which a laser or electron beam fuses powder particles within a bed to form solid layers, as shown in Figure 14 [69, 72, 71].

    **Selective Laser Sintering (SLS)**: SLS is a well-known 3D printing technology renowned for creating complex, functional parts with high accuracy, utilizing a variety of materials [71]. Unlike many additive methods, SLS employs a laser to fuse powdered materials without the need for support structures, offering exceptional design flexibility. In this powder bed fusion process, the powder bed is scanned with a laser, forming solid layers, and the process repeats until the object is complete. The key steps include powder filling, protective gas introduction, and laser scanning. SLS can be classified into direct and indirect methods based on the printer mechanism used [71].

SLS is particularly valued for its ability to create isotropic parts with precise layer binding, enabling intricate designs and faster production compared to conventional methods. Despite its advantages in rapid prototyping and manufacturing functional polymer and ceramic components, SLS does face challenges such as porosity, shrinkage, and the need for post-processing to enhance surface quality [71]. Applications span industries such as aerospace, where non-critical components are produced, and healthcare, where personalized medicines have been developed using pharmaceutical powders. Emerging trends in SLS include creating flexible electronics with metal nanoparticles for micropatterning, with AI increasingly utilized to reduce defects like gas bubbles [71].

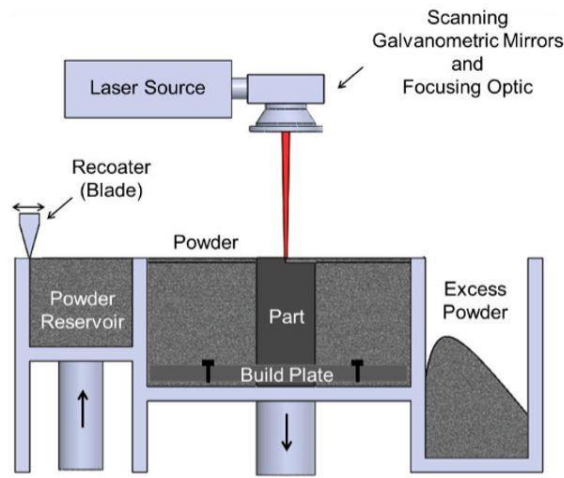

Figure 14: Schematic representation of the PBF additive manufacturing process [75]

**Selective Laser Melting (SLM)**: Also referred to as Laser Powder Bed Fusion (LPBF), SLM is an advanced, high-precision AM technique primarily used for metals [71]. The process involves a high-powered laser that fully melts and fuses layers of metallic powder deposited on the build bed. This starts with an even powder distribution, followed by controlled laser heating of the powder bed above the melting point to form a melt pool. The molten material quickly cools and solidifies, forming fused tracks layer-by-layer as the build platform lowers incrementally. Post-process recovery ensures up to 99% recycling of unused powder, enhancing material efficiency [71].

SLM is notable for producing high-density, net-shaped parts with exceptional mechanical properties, achieving near-full density (up to 100%) and superior tensile strength. The typical layer thickness ranges from 20 to 100 $\mu$m, with particle sizes ranging between 20–50 $\mu$m, enabling intricate designs with tolerances of ±0.2%. The process is predominantly used with metals like titanium alloys (e.g., Ti6Al4V), stainless steels (316L, 304), aluminum alloys (AlSi10Mg), and cobalt-chrome alloys. However, challenges such as residual stress, distortion, and impurities due to high temperatures (often exceeding 1600°C) remain. Support structures are necessary to mitigate these issues, while inert gas environments prevent oxidation during the printing process. Post-processing, including CNC machining and surface cleaning, further enhances the final product quality [71].

Applications include aerospace, automotive, medical devices, and patient-specific implants. Notable advancements include the use of high-entropy alloys, functionally graded materials, and optimization through AI and machine learning to improve mechanical properties. Additionally, multi-laser systems aim to enhance production efficiency, and in-situ monitoring ensures consistent quality during the printing process. Sustainability efforts focus on powder recycling and improving energy efficiency [71].

**Sheet Lamination**: Sheet Lamination encompasses techniques like Laminated Object Manufacturing (LOM) and Ultrasonic Additive Manufacturing (UAM), where thin layers of material are bonded to form 3D parts [72, 76]. LOM uses readily available paper sheets, bonded with adhesive, making it an affordable option for producing visual models. Cross-hatching within the design aids manual removal during post-processing, though parts typically require finishing techniques such as painting or sanding for enhanced aesthetics. UAM, on the other hand, uses metallic sheets (e.g., aluminum or titanium), bonded via ultrasonic welding at low temperatures [72, 76]. This technique allows for the creation of internal geometries and the embedding of electronics, with plastic deformation enhancing interlayer bonding. CNC machining is needed to remove unbound material, ensuring high material efficiency with up to 99% recyclability. Both LOM and UAM are costeffective, fast, and easy to handle, though they face challenges such as limited material options and the need for post-processing to improve surface finishes. LOM is ideal for prototypes, while UAM shows promise for functional multi-material applications, highlighting the versatility of sheet lamination techniques [72, 69].

**Directed Energy Deposition (DED)**: DED is an advanced AM technology that utilizes focused thermal energy such as a laser, electron beam, or plasma arc to melt and deposit material (typically metal powder or wire), layer by layer [71]. Controlled by CAD files, DED enables high-quality part production with material efficiency, design flexibility, and the ability to repair or augment existing components. The process works by delivering material through a multi-axis nozzle, where thermal energy melts the material during deposition, creating fully dense parts with strong bonds to the substrate. While precise, DED often requires post-processing to refine surface finishes. Key advantages include minimal material waste, the ability to create complex geometries, and compatibility with difficult-to-process alloys, making it suitable for large components. However, limitations include rough surface finishes, lower precision compared to other AM methods, and the need for post-processing to achieve finer details [71].

DED supports a wide range of materials, including stainless steel, titanium alloys, nickel-based alloys, aluminum, cobalt–chrome, ceramics, and composites. Precursor powders must meet stringent requirements for particle size, flowability, and purity to optimize deposition and material properties. Ongoing advancements aim to expand material compatibility, further enhancing industrial applications [71].

Applications of DED span aerospace (e.g., turbine blades, structural repairs), energy (components for oil, gas, and power generation), and healthcare (customized implants and prosthetics). The technology also aids tooling applications by adding features to molds and dies. Research in prototyping and material testing continues to push the boundaries of DED's capabilities [71].

Emerging trends in DED include hybrid manufacturing, combining additive and subtractive processes, multi-material printing for functionally graded designs, and real-time monitoring systems to ensure part quality. Advanced software optimizations and sustainability efforts (e.g., powder recycling, energy efficiency) are also central to the continued evolution of DED, positioning it as a transformative technology across multiple sectors [71].

### 3.2. Additive Manufacturing in the Space Industry

The integration of AM into space systems commenced with the first deployment of a 3D printer aboard the International Space Station (ISS) in 2014, under NASA's In-Space Manufacturing initiative [77]. This effort was followed by NASA's Additive Manufacturing Facility (AMF), created by Made In Space, Inc., which has fabricated functional items such as tools and brackets, with dimensions up to 14×10×10 cm, aboard the ISS [78]. Similarly, the European Space Agency (ESA) has advanced AM capabilities, recently deploying a metal 3D printer on the ISS for manufacturing metallic components in orbit [79].

AM plays a pivotal role in space applications by minimizing mass and optimizing structural designs. Coupled with topology optimization techniques, it can achieve up to 70% mass reduction in components [80]. The ability to manufacture on-demand in orbit reduces reliance on resupply missions and enhances mission flexibility, a critical aspect for prolonged deep-space missions [77, 79].

Notable projects include ESA's IMPERIAL initiative, which enables the continuous printing of large polymer structures, and the RegISS program, which explores the use of lunar regolith composites for in-situ resource utilization on planetary surfaces [81, 82].

A significant advantage of AM is its capacity to reduce part mass without compromising structural integrity. The integration of topology optimization during the design phase allows the removal of non-load-bearing material, achieving up to 70% weight savings compared to traditional manufacturing methods [83, 71]. This reduction in mass is crucial for minimizing launch costs and enhancing mission payload capabilities.

Beyond weight reductions, AM facilitates the consolidation of multi-part assemblies into monolithic structures, eliminating mechanical interfaces prone to fatigue and failure. This integration improves overall reliability and enables the incorporation of internal features, such as fluid channels and lattice structures, that are either unfeasible or cost-prohibitive with conventional methods [71].

Moreover, AM processes, such as powder-bed fusion and filament deposition, ensure near-net-shape fabrication, drastically reducing material waste. Scrap rates drop by over 90% compared to subtractive machining, aligning with sustainability objectives in both terrestrial and space manufacturing [71].

Perhaps most transformative is the capability of AM for in-orbit manufacturing. NASA's ISM printer, demonstrated aboard the ISS in 2016, successfully produced missioncritical tools and replacement parts in microgravity, reducing dependency on Earthbased resupply and expediting repair processes [77]. The subsequent deployment of the AMF expanded the range of printable items, including brackets and wrenches made of polymers up to 14×10×10 cm [78]. ESA's addition of a metal 3D printer on the ISS has extended these capabilities to include high-temperature alloys, enabling the production of metallic components for structural and propulsion systems [79]. These advancements collectively bolster mission resilience and adaptability, essential for sustainable, longduration exploration beyond LEO.

*3.3. Challenges in Metallic Additive Manufacturing*

Despite the transformative potential of metallic AM, particularly Laser Powder Bed Fusion (LPBF), several technical and regulatory hurdles continue to impede its widespread adoption for critical space hardware. One of the most pervasive issues is internal porosity and lack of full melt fusion, where insufficient energy density or suboptimal scan strategies leave voids that act as crack initiation sites under cyclic loading, significantly reducing fatigue life and overall mechanical reliability [84, 71]. Moreover, the as-built surface roughness inherent to powder-bed processes often exceeds the stringent finish requirements of aerospace components, necessitating costly and time-consuming post-processing steps such as machining, shot-peening, or chemical polishing to achieve acceptable tolerances [84]. Rapid thermal cycling during layer deposition introduces substantial residual stresses and thermal gradients, leading to part distortion, delamination, or even catastrophic cracking if not carefully managed through optimized support structures and scan path algorithms [71].

Compounding these process-related defects, the quality and consistency of the metal powder feedstocks, defined by the particle size distribution, morphology, flowability, and chemistry, critically influence build integrity; however, standardized, aerospace-grade metal powders remain scarce and expensive, and variations between suppliers can produce unpredictable properties in the final part [85]. On the regulatory front, flight-qualification of AM parts demands exhaustive testing, including non-destructive evaluation (NDE) of complex internal geometries and adherence to rigorous standards such as NASA-STD-6030, making certification more time- and resource-intensive than for conventionally manufactured components [86].

To address these challenges, researchers are pursuing advanced in situ monitoring techniques, such as high-speedd infrared thermography and optical coherence tomography to detect defects layer by layer, paired with machine learning-drivenn control systems that adjust process parameters in real time to mitigate pore formation and thermal stress accumulation [87]. Concurrently, the development of tailored alloy compositions and powder conditioning methods aims to improve powder flowability and enhance melt pool stability, further reducing defect prevalence and moving metallic AM closer to fulfilling the stringent demands of future deep-space missions [87].

## 4. Discussion

The proliferation of satellite constellations in LEO has fundamentally altered the risk landscape for spacecraft, exposing vehicles to a dual threat from naturally occurring micrometeoroids and growing concentrations of anthropogenic debris. While micrometeoroids permeate all orbital regimes, particularly below 270 km and above 4800 km, debris is heavily concentrated between 600 km and 1300 km, where recently launched constellations now operate [38, 10]. Even sub-millimeter particles, traveling at velocities up to 70 km/s, can initiate damage mechanisms, such as solar array degradation or structural pitting that cumulatively reduce satellite lifetimes by up to 60 % [34, 2].

Active mitigation strategies (e.g., conjunction assessment-driven maneuvers or endof-life deorbit kits) have demonstrably lowered collision probabilities for large, trackable debris, but remain powerless against the vast reservoir of small, untrackable fragments and micrometeoroids [41, 43]. Thus, passive shielding persists as the primary defense. Classical Whipple and multi-shock shields employ sacrificial bumpers to fragment incoming projectiles, followed by intermediate fabrics or foams to disperse fragment clouds and a rear wall to arrest residual energy [16, 45, 46]. Enhanced configurations, such as Whipple shields stuffed with Kevlar/Nextel interlayers or mesh double bumpers, demonstrate significant gains in mass-specific protection and have been validated on platforms such as the ISS and Space Shuttle [57, 46].

Additive manufacturing (AM) introduces new design freedoms for MMOD protection. Powder-bed fusion (SLS/SLM) enables graded-porosity lattices, while directed energy deposition and sheet lamination support repair-on-orbit and multi-material integration of shielding features [69, 71]. On-station AM demonstrations have confirmed the feasibility of producing polymer tools and small structural parts in microgravity, heralding a shift toward responsive, in-situ fabrication of bespoke shield geometries [77, 79]. Yet, to realize AM's full potential for critical hardware, challenges in internal porosity, residual stress, and surface finish must be addressed through in-situ monitoring, and machine learning–driven control systems, alongside rigorous flight qualification protocols (e.g., NASA-STD-6030) [84, 86, 87].

Material selection standards from flammability and outgassing (ASTM E-595) to plasma-charging resistance (NASA-HDBK-4002) and ESA's PSS-01-701 ensure that shield constituents meet the multifaceted demands of the space environment, including thermal control, structural integrity, and crew safety [62, 63, 64]. The integration of high-Z fillers into UHMWPE/epoxy composites further extends functionality to include gamma-ray attenuation without prohibitive mass increases [88].

## 5. Outlook

Lightweight shielding solutions capable of withstanding hypervelocity impacts (HVI) remain critical for spacecraft design. Porous metallic lattice structures combined with advanced polymer sheets present a promising approach for energy absorption and spacecraft protection. Building on the demonstrated performance of open metallic foams with bumper plates [89], we propose a hybrid configuration featuring a porous metallic lattice coupled with a thin polymer sheet composed of epoxy resin reinforced with ultra-high molecular weight polyethylene (UHMWPE).

Epoxy resins offer excellent adhesion, chemical resistance, and processability as matrix materials. When combined with UHMWPE fibers or granules, these composites leverage the exceptional tensile strength and low density of polyethylene while maintaining epoxy's stiffness and durability. Recent advances in surface modification techniques including plasma treatment, chemical etching, and grafting have improved fiber–matrix adhesion by over 50%, significantly enhancing interfacial shear strength and impact toughness [90].

The functionality of UHMWPE-epoxy sheets can be further augmented through hybridization with natural graphite (NG) flakes. Using a mixed-heating powder method, NG aligns horizontally on UHMWPE granules, yielding composites with anisotropic thermal conductivity (19.87 W/(m·K) in-plane; 10.67 W/(m·K) through plane at 21.6 vol% NG) [91]. This represents nearly two orders of magnitude improvement over neat epoxy while matching certain lightweight metals, providing significant thermal management capability during MMOD impacts or radiation exposure.

Radiation-shielding performance can be enhanced through incorporation of high-Z fillers. NASA Langley research demonstrates that UHMWPE-based systems processed via field-assisted sintering technology (FAST) with boron carbide, tungsten carbide, or gadolinium achieve near-theoretical densities and maintain structural integrity under radiation [88]. Gamma spectroscopy reveals up to 40% attenuation relative to aluminum without proportional mass penalties [88].

When deployed as discrete sheets in conjunction with metallic lattices, these hybrid UHMWPE/NG-epoxy composites offer a multifunctional shielding solution:

- Metallic lattices provide geometric stiffness and primary fragmentation control

- Polymer sheets contribute impact energy dissipation through controlled deformation

- Anisotropic thermal pathways enable efficient heat management

- Radiation attenuation complements micrometeoroid protection

This configuration addresses the concurrent challenges of micrometeoroid impacts, thermal fluxes, and radiation within mass-constrained spacecraft systems, representing a viable path toward next-generation multifunctional shielding architectures.

## 6. Conclusion

This work underscores that protecting future space assets, and, by extension, human explorers, requires a holistic approach that combines active avoidance, stringent material standards, and advanced passive shielding solutions. Multilayered architectures, from classical Whipple shields to next-generation hybrid configurations, remain essential to withstand the unpredictable mix of micrometeoroids and orbital debris. Additive manufacturing offers a paradigm shift, enabling on-demand, topology-optimized, multifunctional shield components, but its transition to flight-qualified hardware hinges on overcoming powder-related defects and securing exhaustive certification pathways.

The proposed outlook of porous metallic lattices paired with polymer like UHMWPE/graphite-enhanced epoxy sheets presents a promising route toward massefficient multifunctional protection, simultaneously addressing impact energy dissipation, thermal management, and radiation attenuation. Continued research into surface-modified fiber-matrix adhesion, anisotropic thermal composites, and field-assisted sintering of high-Z fillers will further refine these hybrid shields.

Ultimately, preserving the sustainability of near-Earth orbital regimes, and ensuring the safety of future cislunar and deep-space missions will depend on international collaboration to enforce debris mitigation policies, advance active debris removal, and adopt innovative materials and manufacturing techniques. By integrating these strategies, we can chart a viable path toward resilient, adaptable spacecraft that safeguard both mission success and the orbital environment for generations to come.